\documentstyle[12pt]{article}
\def\R{{\bf R}}

\begin{document}

\title {\large \bf Features of the Extension of a Statistical
Measure of Complexity to Continuous Systems}
\author{\normalsize Raquel G. Catal\'an$^{*}$, Jos\'e Garay$^{\dag}$ and
Ricardo L\'{o}pez-Ruiz$^{\ddag}$ \\
                                   \\
\small $^{*}$ Dpto. de Matem\'aticas e Inform\'atica,\\
 \small Universidad P\'ublica de Navarra, $31008$-Pamplona, Spain.\\
 \small $^{\dag}$ Departamento de Matem\'aticas, \\
 \small $^{\ddag}$ DIIS - \'Area de Ciencias de la Computaci\'on, \\
 \small Facultad de Ciencias, Edificio B, \\
\small Universidad de Zaragoza, $50009$-Zaragoza, Spain.\\
\small {\bf date:} October-2001 }
\date{}

\maketitle
\baselineskip 8mm

{\bf Abstract:} We discuss some aspects of the extension to
continuous systems of a statistical measure of complexity 
introduced by L\'opez-Ruiz, Mancini and Calbet (LMC) [Phys. Lett.
A 209 (1995) 321]. In general, the extension of a magnitude from
the discrete to the continuous case is not a trivial process and
requires some choice. In the present study, several
possibilities appear available. One of them is examined in detail.
Some interesting properties desirable for any magnitude of
complexity are discovered on this particular extension.\newline
{\small {\bf Keywords:} Complexity; Entropy; Disequilibrium; Continuous systems; 
 Invariance properties.}
 {\small \hbox{   }{\bf e-mail}: rilopez@posta.unizar.es}

\newpage
\section{Introduction}

In the last years many {\it complexity measures} have been
proposed as indicators of the complex behaviour found in different
systems scattered in a broad spectrum of fields. Some of them come
from physics such as the effective measure of complexity
\cite{grassberger}, the thermodynamical depth \cite{lloyd} and the
simple measure of complexity \cite{shiner}. Other attempts arise
from the field of computational sciences such as algorithmic
complexity \cite{kolmogorov, chaitin}, Lempel-Ziv complexity
\cite{lempel} and $\epsilon$-machine complexity
\cite{crutchfield}. Other works try to enlighten this question in
many other contexts: ecology, genetics, economy, etc., for
instance, the complexity of a system based on its diversity
\cite{huberman} and the physical complexity of genomes
\cite{adami}.

Most of these proposals coincide in using concepts such as entropy
(in physics) or information (in computational sciences) as a basic
ingredient for quantifying the complexity of a phenomenon. Also
there is a general belief that the notion of complexity in physics
must start by considering the perfect crystal and the isolated
ideal gas as examples of simple models with zero complexity. Both
systems are extrema in an {\it entropy/information scale} and
therefore some fundamental ingredient would be missing if one
insisted on describing complexity only with the ordinary
information or entropy.

It seems reasonable to adopt some kind of distance to the
equipartition, the disequilibrium of the system, as a new
ingredient for defining an indicator of complexity. Going back to
the two former examples, it is readily seen that they are extremes
in a {\it disequilibrium scale} and therefore disequilibrium
cannot be directly associated with complexity.

The recently introduced L\'{o}pez-Ruiz-Mancini-Calbet (LMC)
statistical measure of complexity \cite{lopez} identifies the
entropy or information stored in a system and its distance to the
equilibrium probability distribution (the disequilibrium) as the
two ingredients giving the correct asymptotic properties of a
well-behaved measure of complexity. In fact, it vanishes both for
completely ordered and for completely random systems. Besides
giving the main features of an intuitive notion of complexity, it
has been shown that LMC complexity successfully enables us to
discern situations regarded as complex in discrete systems out of
equilibrium: one instance of a local transition to chaos via
intermittency in the logistic map \cite{lopez}, the dynamical
behaviour of this quantity in a simplified isolated gas
\cite{lopez1} and another example of classical statistical
mechanics \cite{lopez2}.

A possible formula of LMC complexity for continuous systems was
suggested in L\'opez-Ruiz {\it et al.} \cite{lopez}. Anteneodo and
Plastino \cite{plastino} pointed out some peculiarities concerning
such an extension for continuous probability distributions. 
It is the aim of this work to offer a discussion of the
extension of LMC complexity for continuous systems. 
A slightly modified extension brings
interesting and very striking properties, and some of the Anteneodo
and Plastino remarks are solved with the new proposed definition.

In Section 2 the extension of information and disequilibrium
concepts for the continuous case are discussed. In Section 3 the
LMC measure of complexity is rewieved and possible extensions for
continuous systems are suggested. We proceed to present some
properties of one of these extensions in Section 4. Finally,
we establish the Conclusions.

\section{Entropy/Information and Disequilibrium}
\label{sec:h-d}

Depending on the necessary conditions to fulfill, the extension of
an established formula from the discrete to the continuous case
always requires a careful study and in many situations some kind
of choice between several possibilities. Next we carry out this
process for the entropy and disequilibrium formulas.

\subsection{Entropy or Information}
\label{sec:h}

Given a discrete probability distribution $\{p_i\}_{i=1,2,...,N}$
satisfying $p_i\geq 0$ and $\sum_{i=1}^N p_i =1$, the {\it
Boltzmann-Gibss-Shannon formula} \cite{shannon} that accounts for
the entropy or information, $H$, stored in a system is defined by
\begin{equation}
H(\{p_i\}) = -k \sum_{i=1}^N p_i\log p_i \; , \label{eq:def-h}
\end{equation}
where $k$ is a positive constant. Some properties of this quantity
are: (i) {\it positivity}: $H\geq 0$ for any arbitrary set
$\{p_i\}$, (ii) {\it concavity}: $H$ is concave for arbitrary
$\{p_i\}$ and reaches the extremal value for equiprobability
($p_i=1/N$ $\forall i$), (iii) {\it additivity}: $H(A\cup
B)=H(A)+H(B)$ where $A$ and $B$ are two independent systems, and
(iv) {\it continuity}: $H$ is continuous for each of its arguments.
And vice versa, it has been shown that
the only function of $\{p_i\}$ verifying the latter properties is
given by Eq. (\ref{eq:def-h}) \cite{shannon, khinchin}.
For an isolated system, the {\it irreversibility} property is 
also verified, that is, the time derivative of $H$ is
positive, $dH/dt\geq 0$, reaching the equality
only for equilibrium. 

Calculation of $H$ for a continuous probability distribution
$p(x)$, with support on $[-L,L]$ and $\int_{-L}^{L}p(x)\,dx = 1$,
can be performed by dividing the interval $[-L,L]$ in small
equal-length pieces $\Delta x = x_i-x_{i-1}$, $i=1,\cdots,n$, with
$x_0=-L$ and $x_n=L$, and by considering the approximated discrete
distribution $\{p_i\} = \{p(\bar{x}_i)\Delta x\}$, $i=1,\cdots,n$,
with $\bar{x}_i$ a point in the segment $[x_{i-1},x_i]$. It gives
us
\begin{eqnarray}
H^* & = & H(\{p_i\})\; = \label{eq:def-h*} \\
     & = & -\, k \sum_{i=1}^n
p(\bar{x}_i)\log p(\bar{x}_i)\,\Delta x\, -  \, k \sum_{i=1}^n
p(\bar{x}_i)\log (\Delta x)\,\Delta x . \nonumber
\end{eqnarray}
The second adding term of $H^*$ in the expression
(\ref{eq:def-h*}) grows as $\log n$ when $n$ goes to infinity.
Therefore it seems reasonable to take just the first and finite
adding term of $H^*$ as the extension of $H$ to the continuous
case: $H(p(x))$. It characterizes with a finite number the information
contained in a continuous distribution $p(x)$. In the limit
$n\rightarrow\infty$, we obtain
\begin{eqnarray}
H(p(x)) & = & \lim_{n\rightarrow\infty} \left[-k \sum_{i=1}^n
p(\bar{x}_i)\log p(\bar{x}_i)\,\Delta x \right]\; = \nonumber \\
  & = & -k \int_{-L}^{L} p(x)\log p(x)\,dx . \label{eq:def-hx}
\end{eqnarray}
If $p(x)\geq 1$ in some region, the entropy defined by Eq.
(\ref{eq:def-hx}) can become negative. Although this situation is
mathematically possible and coherent, it is unfounded from a
physical point of view. See \cite{wehrl} for a discussion on this
point. Let $f(p,q)$ be a probability distribution in phase space
with coordinates $(p,q)$, $f\geq 0$ and $dp\,dq$ having the
dimension of an action. In this case the volume element is
$dp\,dq/h$ with $h$ the Planck constant. Suppose that $H(f)<0$.
Because of $\int (dp\,dq/h) f = 1$, the extent of the region where
$f>1$ must be smaller than $h$. Hence a negative classical entropy
arises if one tries to localize a particle in phase space in a
region smaller than $h$, that is, if the uncertainty relation is
violated. In consequence, not every classical probability
distribution can be observed in nature. The condition $H(f)=0$
could give us the mininal width that is physically allowed for the
distribution and so the maximal localization of the system under
study. This {\it cutting} property has been used in the calculations 
performed in Ref. \cite{lopez2}.

\subsection{Disequilibrium}

Given a discrete probability distribution $\{p_i\}_{i=1,2,...,N}$
satisfying $p_i\geq 0$ and $\sum_{i=1}^N p_i =1$, its {\it
Disequilibrium}, $D$, is the quadratic distance of the actual
probability distribution $\{p_i\}$ to equiprobability:
\begin{equation}
D(\{p_i\}) = \sum_{i=1}^N \,\left( p_i - \frac{1}{N} \right)^2 .
\label{eq:def-d}
\end{equation}
$D$ is maximal for fully regular systems and vanishes for
completely random ones.

In the continuous case with support on the interval $[-L,L]$, the
rectangular function $p(x)=1/(2L)$, with $-L<x<L$, is the natural
extension of the equiprobability distribution of the discrete
case. The disequilibrium could be defined as
\begin{displaymath}
D^* = \int_{-L}^L
\left(p(x)-\frac{1}{2L}\right)^2\,dx = \int_{-L}^{L}p^2(x)\,dx -
\frac{1}{2L}\;.
\end{displaymath}
If we redefine $D$ omitting the constant adding term in $D^*$, the
disequilibrium reads now:
\begin{equation}
D(p(x)) = \int_{-L}^L p^2(x)\,dx \;.\label{eq:def-dc}
\end{equation}
$D>0$ for every distribution and it is minimal for the rectangular
function which represents the equipartition. $D$ does also tend
to infinity when the width of $p(x)$ narrows strongly and
becomes extremely peaked.

\section{Statistical Measure of Complexity}

LMC complexity, $C$, has been defined \cite{lopez} as the
interplay between the information, $H$, stored in a system and its
disequilibrium, $D$. Calculation of $C$ for a discrete
distribution $\{p_i\}$, with $p_i\geq 0$ and $i=1,\cdots,N$, is
given by the formula
\begin{eqnarray}
C(\{p_i\}) & = & H(\{p_i\})\cdot D(\{p_i\})\; = \nonumber \\
    & = & -k \left( \sum_{i=1}^N p_i\log p_i \right)\cdot
    \left( \sum_{i=1}^N \,\left( p_i - \frac{1}{N} \right)^2\right)\, .
    \label{eq:def-c}
\end{eqnarray}
This definition fits the intuitive arguments and verifies the
required asymptotic properties: it vanishes for completely ordered
systems and for fully random systems. $C$ has been successfully
calculated in different systems out of equilibrium: one instance
of a local transition to chaos in an unidimensional mapping
\cite{lopez}, the time evolution of $C$ for a simplified model 
of an isolated gas, the 'tetrahedral' gas \cite{lopez1},
some statistical features of the behaviour of LMC
complexity for DNA sequences \cite{zuguo} and a modification of
$C$ as an effective method to identify the complexity
in hydrological systems \cite{gouzhang}.

Feldman and Cruchtfield \cite{feldman} presented as a main
drawback that $C$ vanishes and it is not an extensive variable for
finite-memory regular Markov chains when the system size
increases. This is not the general behaviour of $C$ in the
thermodynamic limit as  it has been suggested by Calbet and
L\'opez-Ruiz \cite{lopez1}. On the one hand, when
$N\rightarrow\infty$ and $k=1/\log N$, LMC complexity is not a
trivial function of the entropy, in the sense that for a given $H$
there exists a range of complexities between $0$ and $C_{max}(H)$,
where $C_{max}$ is given by
\begin{equation}
\left[ C_{max}(H) \right]_{N\rightarrow\infty} =
  H \cdot (1 - H)^2 \; . \label{eq:def-cc1}
\end{equation}
Observe that in this case $H$ is normalized, $0<H<1$, because
$k=1/\log N$. On the other hand, non-extensitivity cannot be
considered as an obstacle since it is nowadays well known that
there exists a variety of physical systems for which the classical
statistical mechanics seems to be inadequate and for which an
alternative non-extensive thermodynamics is being hailed as a
possible basis of a theoretical framework appropriate to deal with
them \cite{tsallis}.

According to the discussion in Section \ref{sec:h-d}, the
expression of $C$ for the case of a continuum number of states,
$x$, with support on the interval $[-L,L]$ and $\int_{-L}^L p(x)\,
dx = 1$, is defined by
\begin{eqnarray}
C(p(x)) & = & H(p(x))\cdot D(p(x))\; = \nonumber \\
    & = & \left( -k \int_{-L}^L p(x)\log p(x)\,dx \right)\cdot
    \left( \int_{-L}^L \,p^2(x)\,dx \right) \; .
    \label{eq:def-cc2}
\end{eqnarray}
Anteneodo and Plastino \cite{plastino} pointed out that $C$ can
become negative. Obviously, $C<0$ implies $H<0$. Although this
situation is coherent from a mathematical point of view, it is not
physically possible. Hence a negative entropy means to localize a
system in phase space into a region smaller than $h$ (Planck
constant) and this would imply to violate the uncertainty
principle (see discussion of Section \ref{sec:h}). Then a
distribution can broaden without any limit but it cannot become
extremely peaked. The condition $H=0$ could indicate the minimal
width that $p(x)$ is allowed to have. Similarly to the discrete
case, $C$ is positive for any situation and vanishes both for an
extreme localization and for the most widely delocalization
embodied by the equiprobability distribution. Thus, LMC complexity
can be straightforwardly calculated for any continuous
distribution by Eq. (\ref{eq:def-cc2}). It has been applied, for
instance, for quantifying  $C$ in a simplified two-level laser
model in Ref. \cite{lopez2}.

Anyway, the positivity of $C$ for every distribution in the continuous
case can be recovered by taking the exponential of $H$. If we
define $\hat{H}=e^{H}$, we obtain a new expression $\hat{C}$ of
the statistical measure of complexity given by
\begin{equation}
\hat{C}(p(x)) =  \hat{H}(p(x))\cdot D(p(x)) = e^{H(p(x))}\cdot
D(p(x))\; .  \label{eq:def-cc3}
\end{equation}
In addition to the positivity, $\hat{C}$ encloses other
interesting properties that we describe in the next section.

\section{Properties of $\hat{C}$}

The quantity $\hat{C}$ given by  Eq. (\ref{eq:def-cc3}) has been
presented as one of the possible extensions of the LMC complexity
for continuous systems. We proceed now to present some of the
properties that characterize such a complexity indicator.

\subsection{Invariance under translations and rescaling
transformations}

If $p(x)$ is a density function defined on the real axis \R,
$\int_{\R} p(x)\,dx = 1$, and $\alpha>0$ and $\beta$
are two real numbers, we denote by $p_{\alpha,\beta}(x)$ the new
probability distribution obtained by the action of a
$\beta$-translation and an $\alpha$-rescaling transformation on
$p(x)$,
\begin{equation}
p_{\alpha,\beta}(x) = \alpha\, p\,(\alpha\, (x-\beta)) \; .
\label{eq:transf1}
\end{equation}
When $\alpha<1$, $p_{\alpha,\beta}(x)$ broadens whereas if
$\alpha>1$ it becomes more peaked. Observe that
$p_{\alpha,\beta}(x)$ is also a density function. After making the
change of variable $y=\alpha (x-\beta)$ we obtain
\begin{displaymath}
\int_{\R} p_{\alpha,\beta}(x)\, dx = \int_{\R} \alpha\,
p\,(\alpha\, (x-\beta))\, dx = \int_{\R} p(y)\, dy = 1\; .
\end{displaymath}
The behaviour of $H$ under the transformation given by Eq.
(\ref{eq:transf1}) is the following:
\begin{eqnarray*}
H(p_{\alpha,\beta}) & = & -\int_{\R} p_{\alpha,\beta}(x) \log
p_{\alpha,\beta}(x)\, dx = -\int_{\R} p(y)\log (\alpha p(y))\, dy
\\
 & = & -\int_{\R} p(y)\log p(y)\, dy - \log \alpha \int_{\R} p(y)\,
 dy \\
 & = & \;\; H(p) - \log \alpha \, .
\end{eqnarray*}
Then,
\begin{displaymath}
\hat{H}(p_{\alpha,\beta}) = e^{H(p_{\alpha,\beta})} =
\frac{\hat{H}(p)}{\alpha} \; .
\end{displaymath}
It is straightforward to see that $D(p_{\alpha,\beta})=\alpha
D(p)$, and to conclude that
\begin{equation}
\hat{C}(p_{\alpha,\beta}) = \hat{H}(p_{\alpha,\beta})\cdot
D(p_{\alpha,\beta}) = \frac{\hat{H}(p)}{\alpha}\, \alpha D(p) =
\hat{C} (p)\; . \label{eq:transf-c-1}
\end{equation}
Observe that translations and rescaling transformations
keep also the shape of the distributions.
Then it could be reasonable to denominate the invariant quantity
$\hat{C}$ as the {\it shape complexity} of the family formed by 
a distribution $p(x)$ and its transformed $p_{\alpha,\beta}(x)$.
Hence, for instance, the rectangular $\Pi (x)$, the isosceles-triangle
shaped $\Lambda (x)$, the gaussian $\Gamma (x)$, or the
exponential $\Xi (x)$ distributions continue to belong to the
same $\Pi$, $\Lambda$, $\Gamma$ or $\Xi$ family, respectively,
after applying the transformations defined by Eq.
(\ref{eq:transf1}). Calculation of $\hat{C}$ on these distribution
families gives us
\begin{eqnarray*}
\hat{C} (\Pi) & = & 1 \\ \hat{C} (\Lambda) & = &
\frac{2}{3}\,\sqrt{e}\approx 1.0991
\\ \hat{C} (\Gamma) & = & \,\sqrt{\frac{e}{2}}\;\approx 1.1658
\\ \hat{C} (\Xi) & = & \;\;\frac{e}{2}\;\;\;\approx 1.3591 \; .
\end{eqnarray*}
Remark that the family of rectangular distributions has a 
smaller $\hat{C}$ than the rest of distributions. This fact is 
true for every distribution and it will be proved in Section
\ref{sec:min-c}.

\subsection{Invariance under replication}

Lloyd and Pagels \cite{lloyd} recommend that a complexity measure
should remain essentially unchanged  under replication. We show
now that $\hat{C}$ is replicant invariant, that is, the shape
complexity of $m$ replicas of a given distribution is equal to the
shape complexity of the original one. \par Suppose $p(x)$ a
compactly supported density function, 
$\int_{-\infty}^{\infty} p(x)\, dx = 1$. Take $n$ copies $p_m(x)$,
$m=1,\cdots,n$, of $p(x)$,
\begin{displaymath}
p_m(x) = \frac{1}{\sqrt{n}}\; p(\sqrt{n}(x-\lambda_m))\, ,\;\;
1\leq m\leq n \, ,
\end{displaymath}
where the supports of all the $p_m(x)$, centered at $\lambda_m's$
points, $m=1,\cdots,n$, are all disjoint. Observe that
$\int_{-\infty}^{\infty} p_m(x)\, dx = \frac{1}{n}$, what make the
union
\begin{displaymath}
q(x)=\sum_{i=1}^n p_m (x)
\end{displaymath}
to be also a normalized probability distribution,
$\int_{-\infty}^{\infty} q(x)\, dx = 1$. For every $p_m(x)$, a
straightforward calculation shows that
\begin{eqnarray*}
H(p_m) & = & \frac{1}{n}\, H(p) + \frac{1}{n} \log\sqrt{n} \\
D(p_m) & = & \frac{1}{n\sqrt{n}}\, D(p) \, .
\end{eqnarray*}
Taking into account that the $m$ replicas are supported on
disjoint intervals on \R, we obtain
\begin{eqnarray*}
H(q) & = & H(p) + \log\sqrt{n}\, , \\
 D(q) & = & \frac{1}{\sqrt{n}}\, D(p) \, .
\end{eqnarray*}
Then,
\begin{equation}
\hat{C}(q) = \hat{C} (p) \, ,
\end{equation}
what completes the proof of the replicant invariance of $\hat{C}$.

\subsection{Near-Continuity}

Continuity is a desirable property of an indicator of complexity.
For a given scale of observation, similar systems should have a
similar complexity. In the continuous case, similarity between
density functions defined on a common support suggests that they
take close values almost everywhere. More strictly speaking, let
$\delta$ be a positive real number. It will be said that two
density functions $f(x)$ and $g(x)$ defined on the interval
$I\in\R$ are {\it $\delta$-neighboring functions} on $I$ if the
Lebesgue measure of the points $x\in I$ verifying $\mid
f(x)-g(x)\mid\geq\delta$ is zero. A real map $T$ defined on
density functions on $I$ will be called {\it near-continuous} if
for any $\epsilon>0$ there exists $\delta(\epsilon)>0$ such that
if $f(x)$ and $g(x)$ are $\delta$-neighboring functions on $I$
then $\mid T(f)-T(g)\mid <\epsilon$.

It can be shown that the information $H$, the disequilibrium $D$
and the shape complexity $\hat{C}$ are near-continuous maps on the
space of density functions defined on a compact support. We must
stress at this point the importance of the compactness condition
of the support in order to have near-continuity. Take, for
instance, the density function defined on the interval $[-1,L]$,
\begin{equation}
g_{\delta,L}(x)= \left\{
\begin{array}{cl}
1-\delta & \mbox{if}\;\; -1\leq x\leq 0 \\
 \frac{\delta}{L} & \mbox{if}\;\;\;\;\;\;\, 0\leq x\leq L \\
 0 & \mbox{otherwise} \; \; ,
\end{array}  \right.
\label{eq:gdl}
\end{equation}
with $0<\delta<1$ and $L>1$. If we calculate $H$ and $D$ for this
distribution we obtain
\begin{eqnarray*}
H(g_{\delta,L}) & = & -(1-\delta) \log (1-\delta) - \delta
\log\left(\frac{\delta}{L}\right) \\
 D(g_{\delta,L}) & = & (1-\delta)^2 + \frac{\delta^2}{L} \, .
\end{eqnarray*}
Consider also the rectangular density function
\begin{equation}
\chi_{[-1,0]}(x)= \left\{
\begin{array}{cl}
1 & \mbox{if}\;\; -1\leq x\leq 0 \\
 0 &  \mbox{otherwise} \; \; .
\end{array}  \right.
\end{equation}
If $0<\delta<\bar{\delta}<1$, $g_{\delta,L}(x)$ and
$\chi_{[-1,0]}(x)$ are $\bar{\delta}$-neighboring functions. When
$\delta\rightarrow 0$, we have that $\lim_{\delta\rightarrow 0}
g_{\delta,L}(x) = \chi_{[-1,0]}(x)$. In this limit process the
support is maintained and near-continuity manifests itself as
following,
\begin{equation}
\left[\lim_{\delta\rightarrow 0} \hat{C} (g_{\delta,L})\right] =
\hat{C}(\chi_{[-1,0]}) = 1 \, .
\end{equation}
But if we allow the support $L$ to become infinitely large, the
compactness condition is not verified and, although
$\lim_{L\rightarrow\infty} g_{\delta,L}(x)$ and $\chi_{[-1,0]}(x)$
are $\bar{\delta}$-neighboring distributions, we have that
\begin{equation}
\left[\left(\lim_{L\rightarrow\infty} \hat{C}
(g_{\delta,L})\right)\rightarrow \infty \right] \neq
\hat{C}(\chi_{[-1,0]}) = 1 \, .
\end{equation}
Then near-continuity in the map $\hat{C}$ is lost due to the
non-compactness of the support when $L\rightarrow\infty$. This
example suggests that the shape complexity $\hat{C}$ is
near-continuous on compact supports and this property will be
rigorously proved elsewhere.

\subsection{The minimal shape complexity}
\label{sec:min-c}

If we calculate $\hat{C}$ on the example given by Eq.
(\ref{eq:gdl}), we can verify that the shape complexity can be as
large as wanted. Take, for instance, $\delta=\frac{1}{2}$. The
measure $\hat{C}$ reads now
\begin{equation}
\hat{C}(g_{\delta=\frac{1}{2},L}) = \frac{1}{2} \sqrt{L}\left( 1 +
\frac{1}{L}\right) \, .
 \label{eq:cd0.5}
\end{equation}
Thus $\hat{C}$ becomes infinitely large after taking the limits
$L\rightarrow 0$ or $L\rightarrow\infty$. Remark that even in the
case $g_{\delta,L}$ has a finite support, $\hat{C}$ is not upper
bounded. The density functions,
$g_{(\delta=\frac{1}{2}),(L\rightarrow 0)}$ and
$g_{(\delta=\frac{1}{2}),(L\rightarrow \infty)}$, of infinitely
increasing complexity have two zones with different probabilities.
In the case $L\rightarrow 0$ there is a narrow zone where
probability rises to infinity and in the case $L\rightarrow\infty$
there exists an increasingly large zone where probability tends to
zero. Both kind of density functions show a similar pattern to
distributions of maximal LMC complexity in the discrete case,
where there is an state of dominating probability and the rest of
states have the same probability.

The minimal $\hat{C}$ given by Eq. (\ref{eq:cd0.5}) is found when
$L=1$, that is, when $g_{\delta,L}$ becomes the rectangular
density function $\chi_{[-1,1]}$. In fact, the value $\hat{C}=1$
is the minimum of possible shape complexities and it is reached
only on the rectangular distributions. We sketch now some steps
that prove this result.

Suppose
\begin{equation}
f = \sum_{k=1}^n \lambda_k \chi_{E_k}
 \label{eq:def-rect}
\end{equation}
to be a density function consisting of several rectangular pieces
$E_k$, $k=1,\cdots,n$, on disjoint intervals. If $\mu_k$ is the
Lebesgue measure of $E_k$, calculation of $\hat{C}$ gives
\begin{displaymath}
\hat{C}(f) = \prod_{k=1}^n
\left(\lambda_k^{-\lambda_k\mu_k}\right)\cdot\left(\sum_{k=1}^n
\lambda_k^2\mu_k\right) \, .
\end{displaymath}
Lagrange multipliers method is used to find the real vector
 $(\mu_1,\mu_2,\cdots,\mu_n ;\newline
\lambda_1,\lambda_2,\cdots,\lambda_n)$ that makes extremal the
quantity $\hat{C}(f)$ under the condition $\sum_{k=1}^n
\lambda_k\mu_k = 1$. This is equivalent to studying the extrema of
$\log\hat{C}(f)$. We define the function
$z(\lambda_k,\mu_k)=\log\hat{C}(f)+ \alpha\left(\sum_{k=1}^n
\lambda_k\mu_k-1\right)$, then
\begin{displaymath}
z(\lambda_k,\mu_k) = -\sum_{k=1}^n \mu_k\lambda_k \log\lambda_k +
\log\left(\sum_{k=1}^n \mu_k\lambda_k^2\right) +
\alpha\left(\sum_{k=1}^n \lambda_k\mu_k-1\right) \, .
\end{displaymath}
Differentiating this expression and making the result equal to
zero we obtain
\begin{eqnarray}
\frac{\partial z(\lambda_k,\mu_k)}{\partial\lambda_k} & = & -\mu_k
\log\lambda_k - \mu_k + \frac{2\lambda_k\mu_k}{\sum_{j=1}^n
\mu_j\lambda_j^2} + \alpha\mu_k = 0 \label{eq:part1} \\
\frac{\partial z(\lambda_k,\mu_k)}{\partial\mu_k} & = & -\lambda_k
\log\lambda_k + \frac{\lambda_k^2}{\sum_{j=1}^n \mu_j\lambda_j^2}
+ \alpha\lambda_k = 0 \label{eq:part2}
\end{eqnarray}
Dividing Eq. (\ref{eq:part1}) by $\mu_k$ and Eq. (\ref{eq:part2})
by $\lambda_k$ we get
\begin{eqnarray*}
\frac{2\lambda_k}{\sum_{j=1}^n \mu_j\lambda_j^2} + \alpha - 1 =
\log\lambda_k \\
 \frac{\lambda_k}{\sum_{j=1}^n \mu_j\lambda_j^2} + \alpha =
 \log\lambda_k\, .
\end{eqnarray*}
Solving these two equations for every $\lambda_k$ we have
\begin{displaymath}
\lambda_k = \sum_{j=1}^n \mu_j\lambda_j^2 \;\;\;\mbox{for all}\;\;
k \, .
\end{displaymath}
Therefore $f$ is a rectangular function taking the same value
$\lambda$ for every interval $E_k$, that is, $f$ is the
rectangular density function
\begin{displaymath}
f = \lambda\cdot\chi_L \;\;\;\mbox{with}\;\;
\lambda=\frac{1}{\sum_{i=1}^{n}\mu_i}=\frac{1}{L} \, ,
\end{displaymath}
where $L$ is the Lebesgue measure of the support.

Then $\hat{C}(f)=1$ is the minimal value for a density function
composed of several rectangular pieces because, as we know for the
example given by Eq. (\ref{eq:cd0.5}), $\hat{C}(f)$ is not upper
bounded for this kind of distributions.

Furthermore, for every compactly supported density function $g$
and for every $\epsilon>0$, it can be shown that near-continuity
of $\hat{C}$ allows to find a $\delta$-neighboring density
function $f$ of the type given by expression (\ref{eq:def-rect})
verifying $\mid\hat{C}(f) - \hat{C}(g)\mid < \epsilon$. The
arbitrariness of the election of $\epsilon$ brings us to conclude
that $\hat{C}(g)\geq 1$ for every probability distribution $g$.
Thus, we can conclude that the minimal value of $\hat{C}$ is $1$
and it is reached only by the rectangular density functions.

\section{Conclusions}

Complexity theory of discrete systems has been equipped with a new
function that not only vanishes for perfectly ordered and
disordered systems but also has resulted helpful in detecting 
complexity in patterns produced by a process. 
Thus LMC complexity has been shown very useful to quantify
complex behaviour in local transitions to chaos in discrete
mappings \cite{lopez}, it has permitted us
to advance the concept of {\it maximum complexity path} 
in the field of systems far from equilibrium \cite{lopez1}, and,
furthermore, a first attempt to quantify complexity in a model of
a two-level laser system was performed in Ref. \cite{lopez2}. 

Another remarkable feature of LMC complexity is the extension
$\hat{C}$ to the continuous case. Results found in the discrete
and the continuous case are consistent: extreme values of
$\hat{C}$ are observed for distributions characterized by a peak
superimposed to a uniform sea. Other merits of this extension have
been studied and explained in the present work.

First, we find that this quantity is invariant under translations
and rescaling transformations. $\hat{C}$ does not change if the
scale of the system is modified but its shape is maintained. It
has been calculated on different families of distributions
invariant under those transformations. The result allows us to
consider $\hat{C}$ as a new parameter that characterizes every
family of distributions.

Second, it seems reasonable and intuitive that the complexity of
$m$ replicas of a given system should be the same to the original
one. We show that $\hat{C}$ embodies this property and it is
invariant under replication.

Third, continuity is not an evident property for such a map
$\hat{C}$. Thus the compactness of the support of the
distributions is an important requirement in order to have similar
complexity for neighboring distributions. This condition has been
strictly established and stood out with an example.

Finally, complexity should be minimal when the system has reached
equipartition. We demonstrate that the minimum of $\hat{C}$ is
found on the rectangular density functions. Its value is
$\hat{C}=1$. Moreover, $\hat{C}$ is not an upper bounded function
and it can become infinitely large.

We believe and we hope that the present discussion
on the extension of LMC complexity to the continuous case may trigger
some practical future considerations in the area of complex
systems theory.

{\bf Acknowledgements} R.G. Catal\'an acknowledges Spanish DGES
for partial financial support (Project PB98-0551).

\end{document}